# Simultaneous Electrical and Optical Readout of Graphene-Coated High Q Silicon Nitride Resonators

V.P. Adiga[1], R. De Alba[2], I.R. Storch[2], P. A. Yu[3], B. Ilic[4], R.A. Barton[4], S. Lee[5], J. Hone[5], P.L. McEuen[6], J.M. Parpia[2], H.G. Craighead[1]

1. School of Applied and Engineering Physics, Cornell University

2. Department of Physics, Cornell University

3. Department of Chemical Engineering, California Institute of Technology

4. Cornell NanoScale Science & Technology Facility

5. Columbia University

6. Laboratory of Atomic and Solid State Physics and Kavli Institute at Cornell for Nanoscale Science

**We have fabricated and tested mechanical resonators consisting of a single-atomic-layer of graphene deposited on suspended silicon nitride membranes. With the addition of the graphene layer we retain the desirable mechanical properties of silicon nitride but utilize the electrical and optical properties of graphene to transduce resonant motion by both optical and electrical means. By positioning the graphene-on-silicon-nitride drums in a tunable optical cavity we observe position dependent damping and resonant frequency control of the devices due to optical absorption by graphene.**

Resonant electromechanical systems[1,2] and optomechanical systems[3] with high quality factors have been studied for applications such as ultrasensitive force measurements and displacement sensing at the quantum limit[3]. They have also found use in accelerometers and gyroscopes[4], and show promise for resonant sensing applications[5,6]. Silicon nitride has desirable mechanical properties for microelectromechanical devices (MEMS) and is relatively simple to fabricate. Ultra-thin mechanical resonators made from silicon nitride have been explored for optomechanics[3], mass sensing[7] and force sensing[8] because of their high mechanical quality factors[9–11] ($Q > 10^6$), low masses and low spring constants[3,11]. Recently it has been shown that membrane Q can be enhanced by the right choice of tensile stress, resonator size, mode shape and optimized fabrication techniques[9,10,12]; quality factors of up to 4.4 million can thus be achieved for a 15 nm thick silicon nitride membrane[9]. Such large area, ultra-thin tensioned membranes are useful as optomechanical elements[3,13] whose mechanical degrees of freedom can be easily controlled using light[3,11,13]. However, because of the insulating nature of silicon nitride, some of the most desirable characteristics of these high-Q resonators can only be transduced optically. Electrical integration of these devices can be achieved through deposition of a thin conducting layer on the resonator surface. For metals, however, the thickness required to form a continuous layer results in significantly degraded Q and increased mass[14–16]. Metallization also adds complexity in terms of stresses associated with thermal expansion mismatch, causing the freestanding structures to bend or buckle.

Graphene has been widely studied because of its unique electronic[17], optical[18] and mechanical properties[19]. Its light mass and strong optical absorption make it an ideal candidate for achieving optomechanical coupling[13]. Mechanical resonant devices have been constructed of monolayer graphene[20–24], but the mechanical quality factor, fabrication yield and durability of these structures is limited. Hybrid silicon nitride-graphene (SiNG) devices that combine the properties of both materials would greatly expand the range of possible device applications, combining the desirable mechanical properties of silicon nitride with the electrical and optical properties of graphene. In this article, we demonstrate the electrical actuation of high stress silicon nitride membranes using monolayer graphene in a tunable Fabry-Perot cavity. We also present simultaneous detection of its resonant motion using both optical and electrical means enabling the comparison of the two detection schemes. Strong optical absorption in the atomic monolayer graphene[18] enables photothermal interaction with the high-Q silicon nitride membrane, with associated frequency and damping tunability due to tension modulation in the nitride. The optical detection scheme results in a better signal to noise ratio except near the points where the cavity reflectivity is close to its turning point. However, optical detection also is responsible for the associated photothermal interaction. The electrical detection of this optical interaction over the entire cavity detuning ($z/\lambda$) range is useful to understand the photothermal processes[13] in these heterostructures; it enables us to decouple the resonant motion modulation due to optical absorption from the position-sensitive optical detection scheme. These frequency-tunable optically and electrically coupled systems have applications including oscillators, filters and sensors[25–27]. Electrical integration of the these high Q devices also enables us to understand mechanical nonlinearities[28] and provide greater scope for quantum control and cooling[29].

Silicon nitride-graphene square drums of side length 100 μm - 400 μm were fabricated using KOH etching of the backside of a silicon wafer. CVD-grown graphene was transferred on top of a wafer containing suspended drums and patterned using optical lithography. Electrical contacts to these resonators were defined by patterning metal leads. These graphene on silicon nitride devices are placed in close proximity to a piezo-controlled metallic mirror that forms a tunable optical cavity. Details of the fabrication and moving mirror setup are provided in the supplemental information. Optical detection involves detecting the change in the reflected laser light as the membrane moves in the low finesse optical cavity ($F \approx 1.2$) formed by the membrane and mirror as shown in Figure 1. The metallic mirror used in this cavity also acts as a conductive electrode which is placed in close proximity to the resonator (< 60 µm), where we apply a bias voltage to actuate and tune the resonance of the SiNG membranes electrostatically under high vacuum conditions (< 2× $10^{-6}$ Torr). A fast photodiode and a network analyzer are used to measure the time-varying component of our reflected optical signal. This detected signal is proportional to both the amplitude of the membrane's motion ($\tilde{z}(\omega)$) and the change in cavity reflectance ($R(z)$) with respect to membrane position $(dR/dz)$[30]. The amplitude of the capacitively-driven membrane motion is given by

$$\tilde{z}(\omega) = -\frac{1}{m_{eff}}\frac{C_g}{d}V_g\tilde{V}_g\frac{1}{\omega_0^2 - \omega^2 + i\omega_0\omega/Q}, \quad (1)$$

where $C_g = \varepsilon_0 A/d$ is the membrane-mirror capacitance, $m_{eff}$ is the membrane's effective mass, and $A$ is the membrane area. $V_g$, $\tilde{V}_g$ are the DC gate voltage and AC drive voltage, respectively. $\omega_0 = 2\pi f_0$ and $\omega = 2\pi f$ are the membrane resonant frequency and the drive frequency. Calculations of $R(z)$ and $dR/dz$ as functions of membrane position are provided in the supplemental information. Our electrical detection scheme involves the capacitive detection of membrane motion, and the observed signal ($\tilde{I}$) is given by

$$\tilde{I} = i\omega C_{tot}\tilde{V}_g - i\omega\frac{\tilde{z}(\omega)}{d}C_g V_g. \quad (2)$$

The first term above corresponds to the total capacitive background ($C_{tot}$), due to the device (~1.5 pf) and all parasitic capacitance (~5 pf). The second term is sensitive to membrane motion. This signal is amplified before being routed to a network analyzer for readout.

Figure 2 shows the typical gate tuning of the resonant frequency, where the composite membrane only shows capacitive softening[31] in the applied DC gate voltage range. At a given gate voltage, both optical and electrical resonant response show a Lorentzian behavior (Figure 2D, E), allowing us to extract the fundamental frequency ($f_0$ = 2.8 MHz), the full width at half maximum power (FWHM) $\Gamma$, and the quality factor ($Q = f_0/\Gamma$) of the device. Electrically and optically detected signals give identical $Q$ and resonant frequency measurements (within fitting errors). For the fundamental mode of a tensioned square drum, the resonant frequency is given by $f_0 = \frac{1}{L}\sqrt{\frac{\sigma}{2\rho}}$, where $\sigma$, $\rho$, $L$ are stress, density and side length of the resonator respectively. This yields a tensile stress of 475 MPa in our 100 $\mu$m membrane. As such, quality factors of thin tensioned membranes scale approximately linearly with the aspect ratio (side length/thickness)[9]. We have measured quality factors of up to 70,000 for a 100 µm square graphene-on-silicon-nitride drum for the fundamental mode. A similar 300 $\mu$m membrane of the same thickness yields Q~250,000 (see supplemental information). Graphene contributes marginally to the observed mechanical damping of these structures[32].

Figure 3 shows the optically and electrically detected resonant frequency response as the optical cavity is detuned by stepping the piezo controlled mirror toward the membrane at a fixed incident laser power (0.2 mW). Periodic variations in the resonant frequency are due to the absorption of optical power by graphene and the resulting change in tensile stress of the nitride membrane. Calculations of the cavity reflectance $R(z)$ and graphene absorption $A(z)$ (Figure 3C) depend on the thickness and refractive index of each optical medium, and are made using a standard transfer matrix approach[33]. Further details are provided in the supplemental information. $A(z)$ in these calculations exceeds the well known value of $\pi\alpha \approx 2.3\%$ due to the cavity effect, and the asymmetric cavity response is caused by reflections within the nitride layer. The slight offset of the nodes in the optically detected signal (corresponding to $dR/dz = 0$) relative to the frequency extremes is indicative of additional losses in the cavity – attributed here to absorption by the Ag mirror. Figure 4A shows the electrically-obtained resonant frequency as a function of mirror position for several values of the incident laser power, with corresponding fits based on the calculated optical absorption of graphene in our system (Figure 3C). Nodal positions in the optical data were used to determine several cavity parameters in these fits based on *dR/dz* (see supplemental information). We observe that the magnitude of the frequency variation (defined as the peak-to-peak frequency shift) scales linearly with incident laser power (shown in Figure 4C).

Both optically and electrically obtained data suggest that the mechanism responsible for resonant frequency shifts in our devices is local heating in the membrane resulting from optical absorption by the graphene. Such heating leads to a lowering of the membrane's tensile stress through thermal expansion of the silicon nitride. In the low optical power limit, this results in a frequency shift that varies with temperature as $\frac{\Delta f}{f_0} = -\frac{E_{SiN}\alpha_{SiN}\Delta T}{2\sigma}$, where $E_{SiN}$ and $\alpha_{SiN}$ are the Young's modulus and thermal expansion coefficient of nitride, and $\Delta T$ is the temperature shift due to optical heating. The numerator in this expression is the change in tension caused by expansion of the nitride, and ignores contributions from graphene contraction since the graphene thickness is small compared to silicon nitride and has minimal affect on the overall mechanics. To lowest order, the temperature rise can be approximated by assuming a circular membrane and solving for the equilibrium heat flow radially outward from the laser spot. Including heat dissipation through both the graphene and the nitride, the steady state temperature difference between the membrane edge and the laser spot is $\Delta T = \frac{P_{abs}}{t_{SiN}k_{SiN} + t_G k_G} \cdot \frac{\ln(L/D)}{2\pi}$. Here $P_{abs}$ is the absorbed optical power and *D* is the laser spot diameter. $t_{SiN}$, $t_G$ are the thicknesses of the two materials, and $k_{SiN}$, $k_G$ are the thermal conductivities (30 W/m-K for nitride and 5x10$^3$ W/m-K for graphene). With a laser spot diameter of ~ 8 μm, graphene absorption of 5% inside the cavity (see Figure 3C), and incident power of 195 μW, we thus expect a temperature rise of ~1.3 K. This results in a maximum frequency variation of $\Delta f$ = -2.7 kHz, which is an overestimate (in magnitude) since we have taken the mean membrane temperature to be that directly at the laser spot. This is, however, in excellent agreement with the measured frequency variation of -2.2 kHz (Figure 4C).

While the optical signal strength exhibits variations primarily due to its dependence on *dR/dz*, the electrical signal amplitude (see supplementary information) shows variations mainly due to changes in the effective damping of the resonator (Figure 4B). Such damping variations resulting from photothermal forces have been observed in tensioned graphene drums[13]. Likewise, photothermal forces and radiation pressure effects on bilayer materials have been studied in the past[22,29,30]. Similar effects are possible in our system, with local bimetallic expansion of the membrane breaking the system symmetry and applying a feedback force in the direction of motion. Such a force would be time delayed by the membrane thermal relaxation time, and would affect both device frequency and damping. Estimates of this time constant[24,34,35] ($\omega \tau$ ~ 2,000) indicate that this effect would play a significant role in damping variations, but would have a negligible effect on the frequency. This model predicts an effective damping[34,35] of $\Gamma_{eff} = \Gamma(1 + Q \frac{\omega \tau}{1+\omega^2 \tau^2} \frac{\nabla F}{K})$, where $K$ is the membrane spring constant and $\nabla F$ is the gradient in the bilayer force (w.r.t. mirror position) experienced by the membrane. The expected damping shift should thus vary as *dA(z)/dz*. However, such a model was found to have systematic deviations from our measured damping shifts (see supplemental). Thus, this is likely not the only mechanism influencing the damping of our devices, and further studies are required to understand the feedback forces in these heterostructures.

We have demonstrated the electrical actuation and detection of high Q silicon nitride membranes using a graphene coating to provide a conductive layer for electrical readout in a tunable Fabry-perot cavity. Optical absorption by graphene in the cavity results in position-dependent modulation of the tension in the silicon nitride, leading to discernible resonant frequency shifts due to the high Q of the silicon nitride/graphene resonator. Damping in silicon nitride/graphene membranes strongly depends on their position in the cavity, indicating a photothermal force on the membrane. The resonant motion of this system can thus be effectively enhanced or dampened at will, and is coupled with resonant frequency control. Resonant frequencies of these high-Q systems can also separately be tuned in situ via DC gate voltage. Graphene on silicon nitride heterostructure systems thus provide for independently varying the mechanical, optical, and electrical degrees of freedom of low-mass, high-Q devices. Integrating the device with a smaller gate distance should enable utilization of the transconductance properties of graphene. Furthermore, improving the quality factor and adjusting $\tau$ via device dimensions will further enhance the photothermal interactions, potentially leading to photothermal self-oscillation of these systems.

Acknowledgements:

We thank the support from Cornell NanoScale Science and Technology Facility. We would like to acknowledge Dave MacNeill, Christopher B Wallin, for useful discussions. We acknowledge financial support from the Cornell Center for Materials Research and NSF grants DMR-0908634, ECCS-1001742, DMR 1120296 and AFOSR MURI, FA9550-09-1-0705.

List of Figures

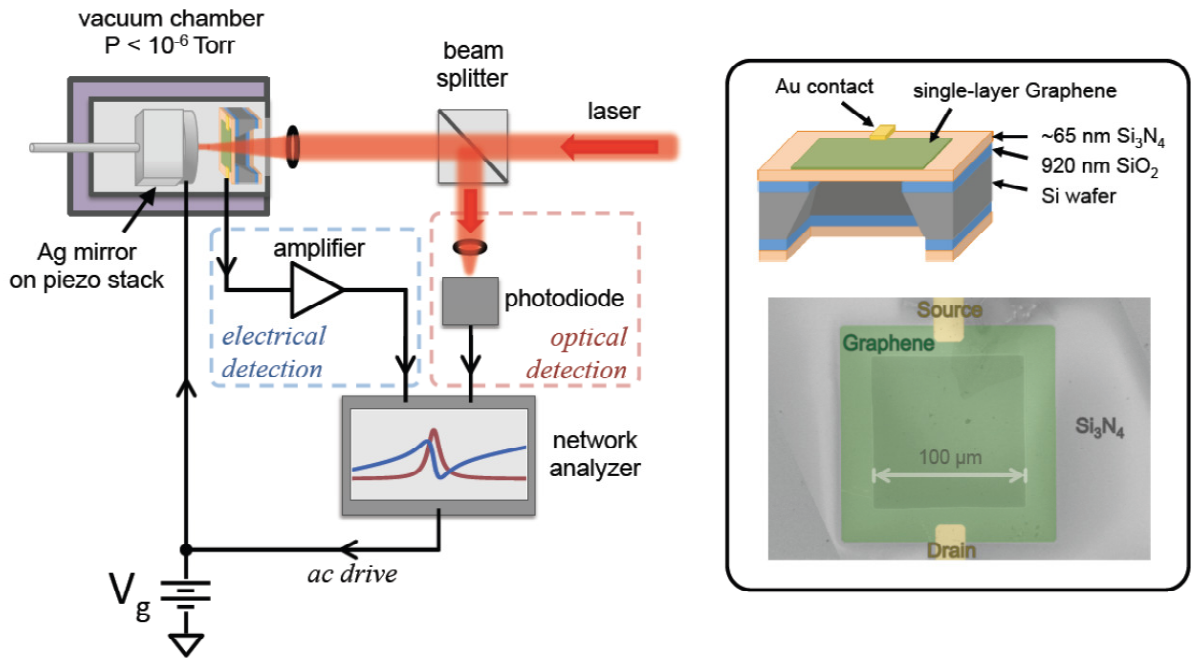

Figure 1: Schematic of the experimental setup. Variation in the reflected light from a Fabry-Perot cavity formed between a graphene on silicon nitride membrane and a piezo-controlled metallic mirror is monitored by a fast photo diode. A gate voltage, $V_g$, is applied between the graphene and the metallic mirror to actuate the resonator; this voltage has a DC component for tuning and an AC component at the drive frequency. Measuring the capacitively coupled current provides a second means to readout mechanical motion. Inset: Combined SEM and optical micrograph of a typical SiNG membrane resonator showing the device layout. Scale bar indicates the suspended region.

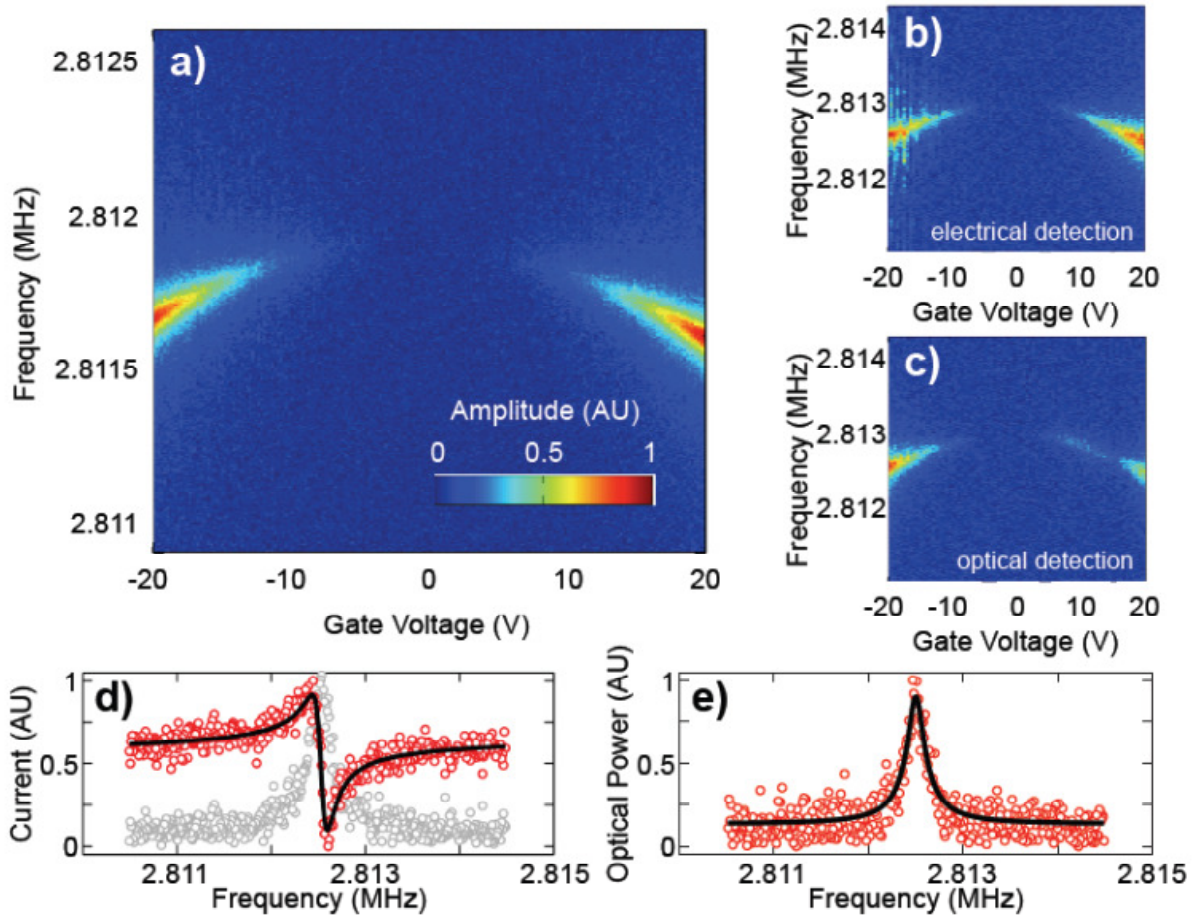

Figure 2: (**a**) Gate tuning resonant frequency of the 100 μm membrane, detected using electrical means with zero laser power. Amplitude is in color scale. The resonator shows capacitive softening in the measured voltage range. Amplitude (detected using both electrical (**b**) and optical (**c**) means at 100 μW laser power) measured as a function of gate voltage. Both electrical and optical detection schemes show reduction in quality factor as a function of gate voltage. (**d**) Sample of raw electrical readout data with fit. Gray data points show the same data with the parasitic capacitance contribution subtracted, illustrating the Lorentzian signal as it appears in the color of (**a**) and (**b**). (**e**) Sample of optical readout data with fit. Fits give $f_0$ = 2.8 MHz, $\Gamma$ = 160 Hz, and Q= 17,000.

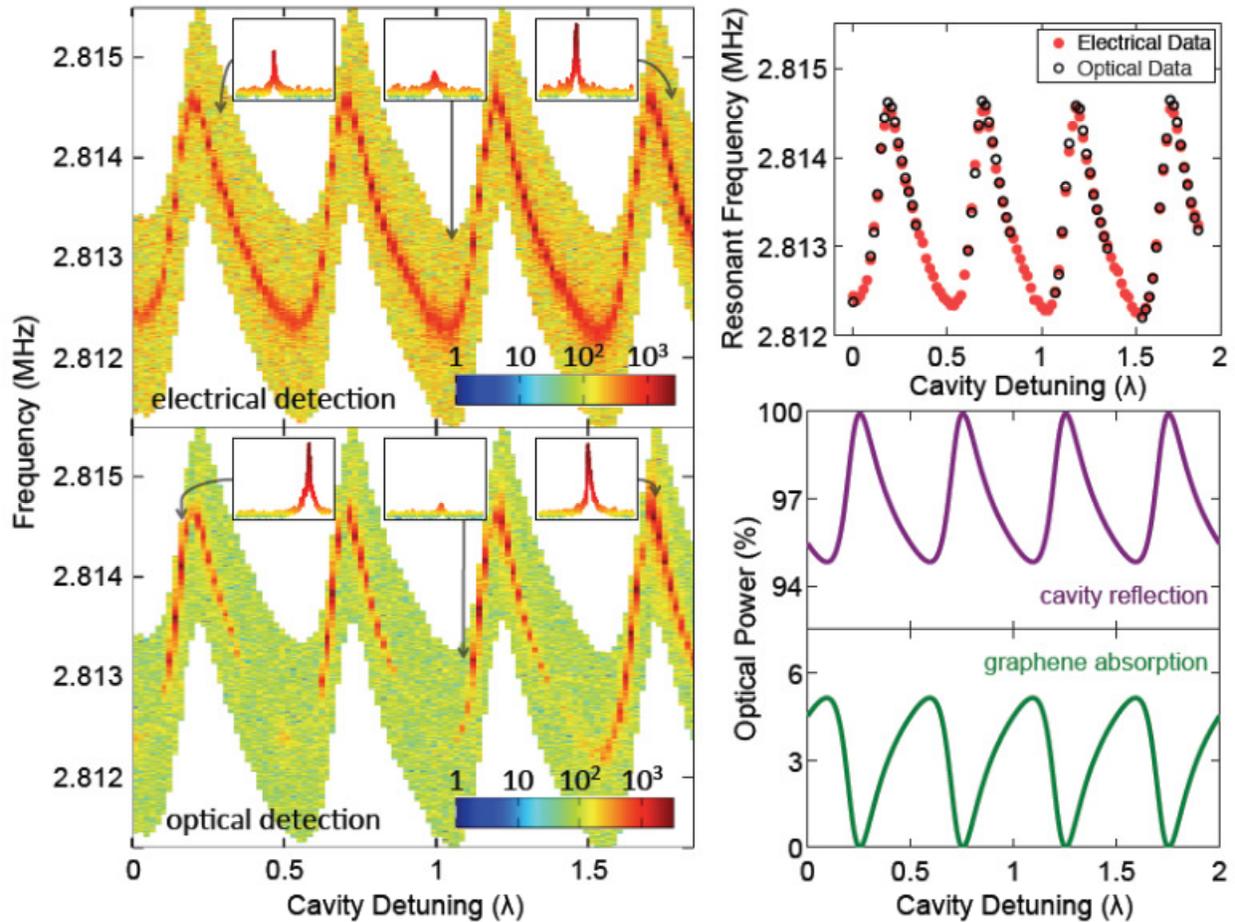

Figure 3: **a)** Electrical and optical detection of resonant frequency as a function of detuning of the cavity by moving the piezo-controlled mirror closer to the membrane. Color scale indicates the amplitude of motion in log scale. The disappearance of the optical readout signal corresponds to the positions in the cavity where *dR/dz* vanishes. The electrically detected signal is continuous and shows an increased signal as the mirror approaches the membrane (capacitive background is subtracted from the data). **b)** Overlaid resonance frequencies from fits of the electrical and optical readout data shown in **(a)**. **c)** Calculated reflectance of the optical cavity (purple) and absorption (green) by graphene as a function of detuning of the cavity.

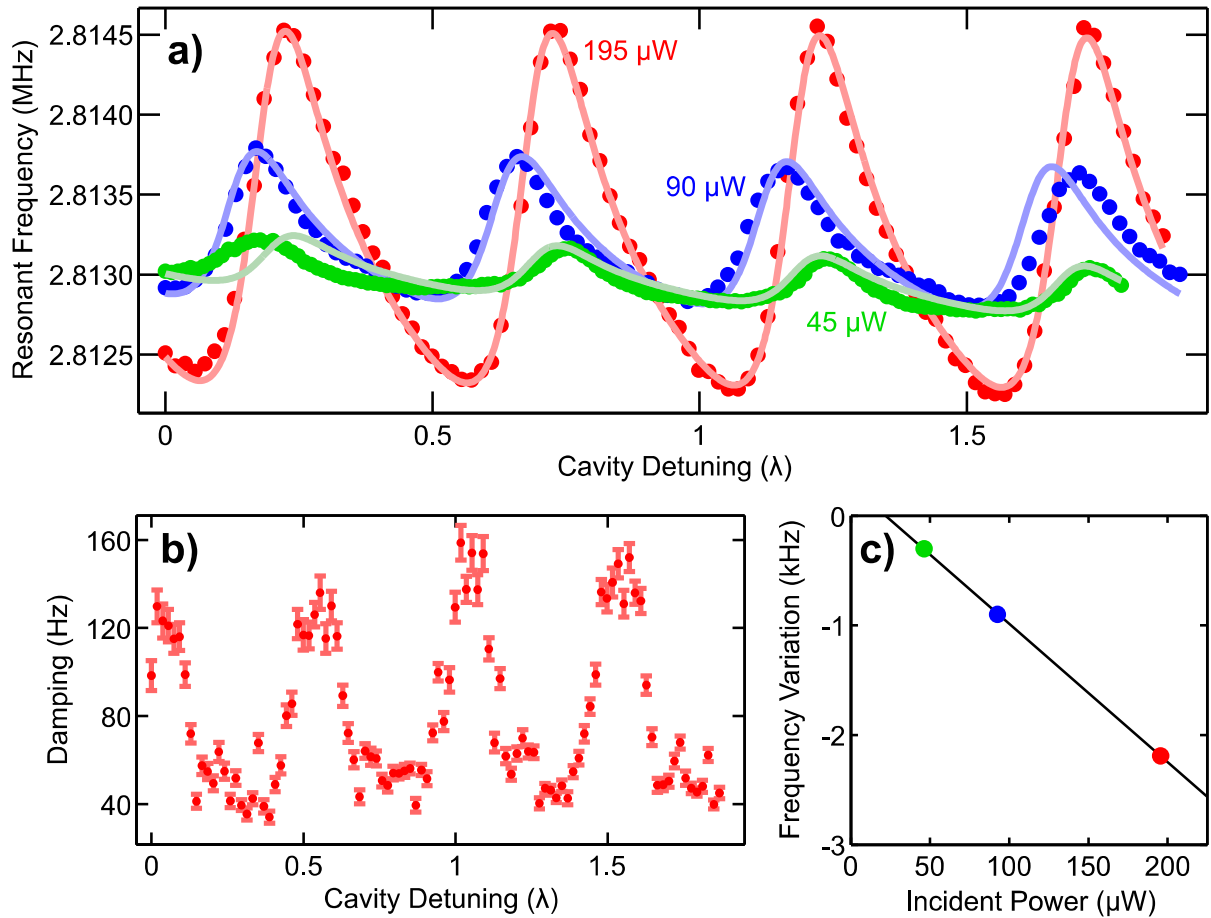

Figure 4: **a)** Resonant frequency of the membrane as a function of incident optical power and cavity detuning measured using electrical detection. Oscillations in the frequency are associated with optical absorption of the graphene, and its effect on the tensile stress of the bilayer membrane. **b)** Measured damping shifts at 195 µW laser power. Damping tuning may be due in part to photothermal feedback on the membrane arising from variations in bilayer expansion. **c)** Maximum frequency variation of the device as a function of incident laser power, with a linear fit.

# Supplemental Information for Simultaneous Electrical and Optical Readout of Graphene coated High Q Silicon Nitride Resonators

V.P. Adiga, R. De Alba, I.R. Storch, P. Yu, B. Ilic, R.A. Barton, S. Lee, J. Hone, P.L. McEuen, J.M. Parpia, H.G. Craighead

**Outline of Supplemental Information:**
1. Fabrication of Graphene on SiN Drums
2. Experimental Setup
3. Electrical Data Fitting
4. Modeling the Optical Cavity
5. Fitting of Cavity Detuning Data
6. Fitting of Damping Shift with Cavity Detuning
7. 300 µm Device Response

## 1. Fabrication of Graphene on Silicon Nitride Drums:

900 nm thick thermal oxide (wet oxide, 980 °C) grown on a double-side-polished wafer (resistance). This oxide provides electrical isolation and etch isolation from a KOH etch. Stoichiometric high-stress nitride (60 nm thick) is grown on thermal oxide at 800 °C. The backside of the wafer is patterned using contact lithography (EV 620) to have square openings. A reactive ion etch ($CHF_3/O_2$ nitride etch chemistry) recipe is used to etch both nitride and oxide. Resist on the backside of the wafer is removed using 1165. Exposed silicon is etched using KOH until the etch stops – when the silicon is completely consumed and oxide interface is reached in the front side of the wafer. Monolayer graphene is grown on copper using a chemical vapor deposition process[1] (980 °C anneal in 60 sccm $H_2$ for 1 hour, graphene growth in 60 sccm $H_2$, 36 sccm $CH_4$ for 30 minutes at 980 °C, followed by cool down to room temperature at 60 sccm $H_2$). CVD-grown graphene is coated with 50 nm of PMMA (1.2 % anisole). Ferric chloride solution is used to etch away the copper, and the graphene is transferred into several DI water baths before transferring onto the final substrate containing square silicon nitride drums on oxide.

Graphene is patterned so that it covers the entire silicon nitride drum (on oxide). The resist and PMMA are removed using 1165 solution. Graphene is annealed at 325 °C for 3 hours using a forming gas mixture ($CH_4$ and Ar, 50 % each at 1 L/min) to remove any residual resist. Gold electrical leads (60 nm thick with 2nm Ti adhesion layer) are patterned on the graphene-coated silicon nitride wafer followed by lift off in 1165. The wafer is spin-coated with SPR 700 to protect the front side, followed by a BOE (6:1) etch for 20 minutes to remove the oxide underneath the silicon nitride. Resist is removed using 1165. Silicon nitride square drums with side length (L) of 100 µm to 400 µm (in increments of 100 µm) are fabricated per die using this method. Samples are diced and each resonator is current annealed to yield at typical resistance of 5 kΩ / square. Source and drain are connected to the co-ax connector using a wire bonder.

## 2. Experimental Setup

Our custom-built tunable cavity setup involves a 3 axis Thorlabs piezo mirror mount (ASM 003) placed on an aluminum base, which is mechanically connected to a micrometer for coarse positioning. The micrometer resolution is 10 µm, whereas the piezo transverse resolution is 10 nm. The total piezo transverse travel is 7 µm.

The piezo mirror mount has both coarse (total travel ~2°) and fine (total travel 2 arc minutes) tilt controls via mechanical screws and 3-axis piezo actuation respectively. This piezo can hold a 7 mm diameter siliver mirror (Thorlabs PF03-03-P01). The mirror is mounted on a custom mirror mount that accepts a coax connector (Molex) and allows the mirror to be very close to the sample (< 50μm). The metallic mirror is electrically connected to a co-ax connector using conductive epoxy.

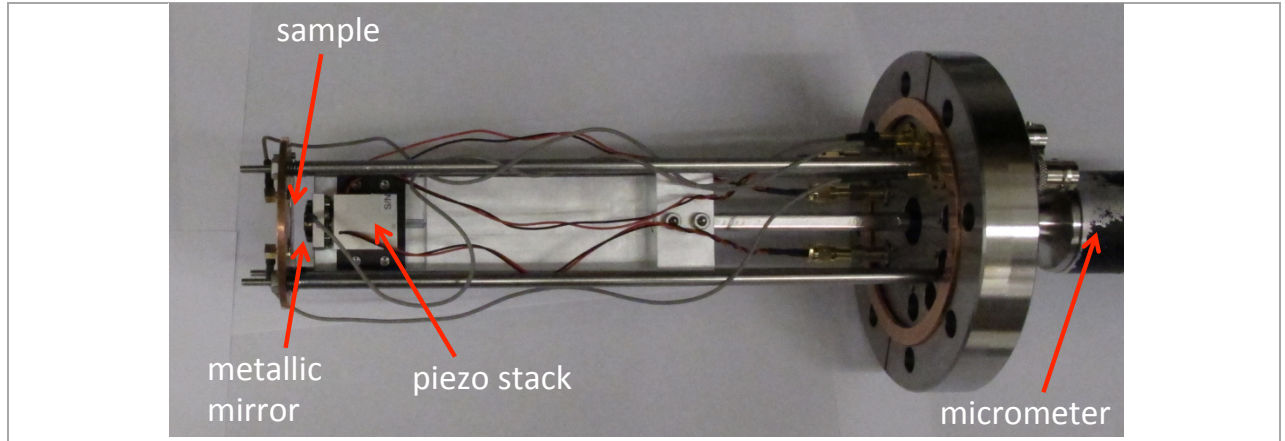

**Figure 1S:** Optical image of the custom-built tunable cavity set up. A micrometer (Huntington Mechanical Laboratories, SN # VF-108) holds an aluminum support piece on which the piezo-controlled mirror is mounted for fine motion. The sample is mounted on an adjustable copper plate supported by steel rods. The whole assembly is mounted on a 4.5-inch flange with external feed-throughs for electrical connection.

The sample is mounted on a copper plate using phosphor bronze springs which result in much reduced mechanical drift. Source and drain are wirebonded to co-ax connectors (Molex) in the copper plate. The copper plate rests on 4 steel rods and can also be tilt-adjusted using set screws. Source, drain and gate are connected to an external feed through a 4.5-inch flange using copper co-ax cables. The sample and mirror are made parallel to each other using optical means before pumping the chamber. Sample is oriented such that graphene faces the metallic mirror. The whole assembly is placed in a vacuum chamber that can reach $2\times10^{-7}$ torr and vacuum is maintained using an ion pump for vibration isolation. A DC+AC bias is applied to the mirror and an AC response through the drain is fed to voltage amplifier followed by a network analyzer. To minimize the parasitic capacitance, membrane resonators (4 per die) along with the electrical leads are positioned such that they are close to the edge of the metallic mirror. Similarly, optical response is read through a fast photo detector using a network analyzer.

### 3. Electrical Data Fitting

In order to obtain reliable mechanical parameters for our resonators, a satisfactory model of the frequency response of our system is needed. As mentioned in the main text, DC and AC voltage biases are applied to the membrane to drive it into mechanical motion. In the case of optical detection, a photodiode captures light reflected from the membrane-mirror cavity, and the measured signal amplitude resembles a standard Lorentzian response in frequency space. For electrical detection, however, the measured signal amplitude is non-Lorentzian. This is due to the many non-resonant components of the current in our circuit. As previously mentioned, the AC current through our device is

$$\tilde{I}(f) = i2\pi f C_{tot} \tilde{V}_g - i2\pi \frac{\tilde{z}(f)}{d} C_g V_g, \quad \text{(Eq 1S)}$$

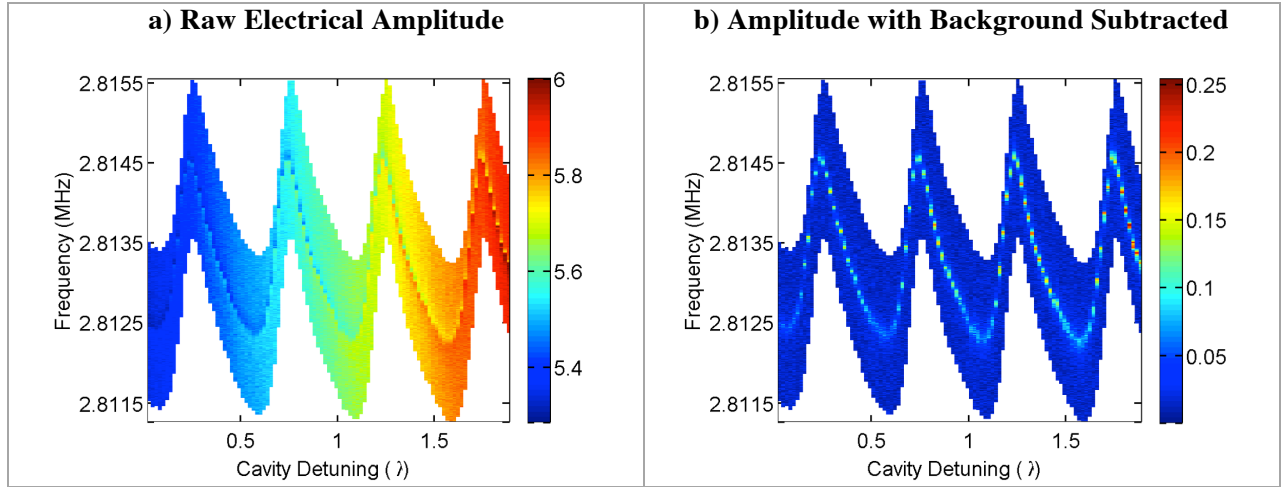

**Figure 2S:** Electrically detected frequency response of our system as the mirror/gate electrode approaches our membrane, with 0.195 mW incident laser power. **a)** Raw amplitude of our measured signal as a function of drive frequency and cavity detuning. Color denotes signal strength in mV. Note the increasing background and resonant signal as the capacitance of the system increases. **b)** Amplitude of the same data (in mV), with background subtraction. Note the uniform background level, and prominent resonance.

where $\tilde{z}(f)$ is the membrane displacement, $d$ is the membrane-mirror distance, and $V_g$, $\tilde{V}_g$ are the DC and AC voltages applied. $C_g$ and $C_{tot}$ are the membrane-mirror capacitance and the total (device + parasitic) capacitance, and $f$ is the drive frequency. Because the linear background changes with the membrane-gate electrode distance, this type of background subtraction is especially useful in analyzing the evolution of our resonance with cavity detuning. Figure 2S shows this evolution as a color plot, before and after subtraction of the linear frequency background.

## 4. Modeling the Optical Cavity

Understanding the distribution of laser light intensity in our system is useful in the interpretation of our cavity detuning measurements. Shifts in the resonant frequency of our device (as seen in Figure 2S) are directly related to the optical power incident on the graphene monolayer, and signal strength in our optically detected data is similarly related to the power reflected from our cavity. We have implemented a standard transfer matrix approach to model our optical system. A schematic of our cavity is shown in Figure 3S.

Routine transfer matrices were applied for each interface and propagation through a homogeneous medium[2]. For clarity, these are:

$$\begin{bmatrix} E_{1+} \\ E_{1-} \end{bmatrix} = \frac{1}{\tau}\begin{bmatrix} 1 & \rho \\ \rho & 1 \end{bmatrix}\begin{bmatrix} E_{2+} \\ E_{2-} \end{bmatrix} \qquad \begin{bmatrix} E_{1+} \\ E_{1-} \end{bmatrix} = \begin{bmatrix} e^{ikl} & 0 \\ 0 & e^{-ikl} \end{bmatrix}\begin{bmatrix} E_{2+} \\ E_{2-} \end{bmatrix} \qquad \text{(Eq 2S)}$$

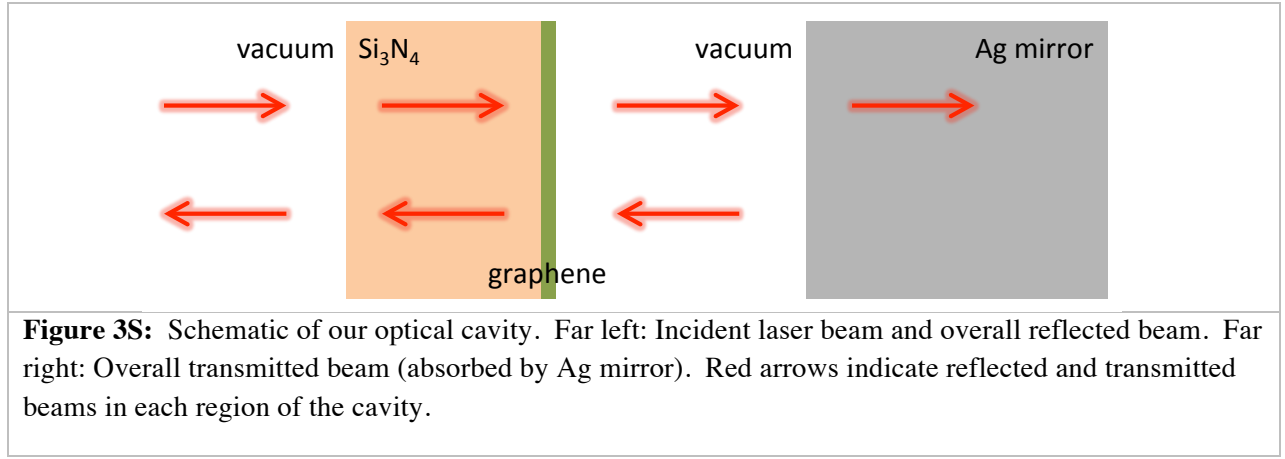

**Figure 3S:** Schematic of our optical cavity. Far left: Incident laser beam and overall reflected beam. Far right: Overall transmitted beam (absorbed by Ag mirror). Red arrows indicate reflected and transmitted beams in each region of the cavity.

where $E_{1+}$, $E_{1-}$ are the right-going and left-going EM waves before an interface (or spatial propagation), and $E_{2+}$, $E_{2-}$ are the corresponding waves after the interface (or spatial propagation). $\rho$, $\tau$ are the interface reflection and transmission coefficients, and $k$, $l$ are the wavenumber and distance traveled. Matrices such as these are applied in succession to find the field in any region of the cavity.

In order to avoid issues with multiple reflected waves within the graphene layer, it was treated as an infinitely thin conducting interface rather than a thin film. To determine the transfer matrix for propagation across the graphene interface, boundary conditions for the $E$ and $B$ fields (governed simply by Maxwell's equations) were used. In short, $E$ fields parallel to the interface are conserved across it, while $B$ fields experience a discontinuity proportional to the free current density of the graphene. Written in matrix form, this is:

$$\begin{bmatrix} E_2 \\ cB_2 \end{bmatrix} = \begin{bmatrix} 1 & 0 \\ -\mu_0 c\sigma & 1 \end{bmatrix} \begin{bmatrix} E_1 \\ cB_1 \end{bmatrix}, \quad \text{(Eq 3S)}$$

where $E_1$, $B_1$ are the total fields before the graphene, and $E_2$, $B_2$ are the total fields after. $\mu_0$, $c$, and $\sigma$ are the vacuum permeability, speed of light in vacuum, and graphene conductivity. Taking $n_1$, $n_2$ to be the refractive indices of the two surrounding media, we can write this relation in terms of the left- and right-going waves as:

$$\begin{bmatrix} E_{1+} \\ E_{1-} \end{bmatrix} = \frac{1}{2n_1} \begin{bmatrix} n_1 + n_2 + \mu_0 c\sigma & n_1 - n_2 + \mu_0 c\sigma \\ n_1 - n_2 - \mu_0 c\sigma & n_1 + n_2 - \mu_0 c\sigma \end{bmatrix} \begin{bmatrix} E_{2+} \\ E_{2-} \end{bmatrix}. \quad \text{(Eq 4S)}$$

If we now use the universal constant $\pi e^2/2h$ for the conductivity of Dirac fermions in graphene [3,4], the combination $\mu_0 c\sigma$ simplifies to $\pi\alpha \approx 0.023$, the well-known opacity of graphene[5]. From Equation 4S, we can thus extract the expected optical transmittance $T_G = (1+\pi\alpha/2)^{-2}$ and reflectance $R_G = \pi^2\alpha^2 T_G/4$ of freestanding graphene.

The relevant optical quantities needed to interpret our data are the optical power absorbed by the graphene membrane $A(z)$ and the total power reflected from our system $R(z)$ (as functions of the membrane-mirror distance). If we assume the mirror to be semi-infinite, we can readily compute the total reflected power and the total power absorbed by the mirror (normalized by the incident laser power) using the matrix approach described above. Of course, these are functions not only of the membrane-mirror distance, but also the $Si_3N_4$ refractive index, $Si_3N_4$ thickness, and the complex refractive index of the mirror. Because the $Si_3N_4$ is considered to be lossless, the remaining optical power that is neither reflected out of the cavity nor absorbed by the mirror can be attributed to absorption by the graphene. Examples of these calculations can be seen in Figure 4S.

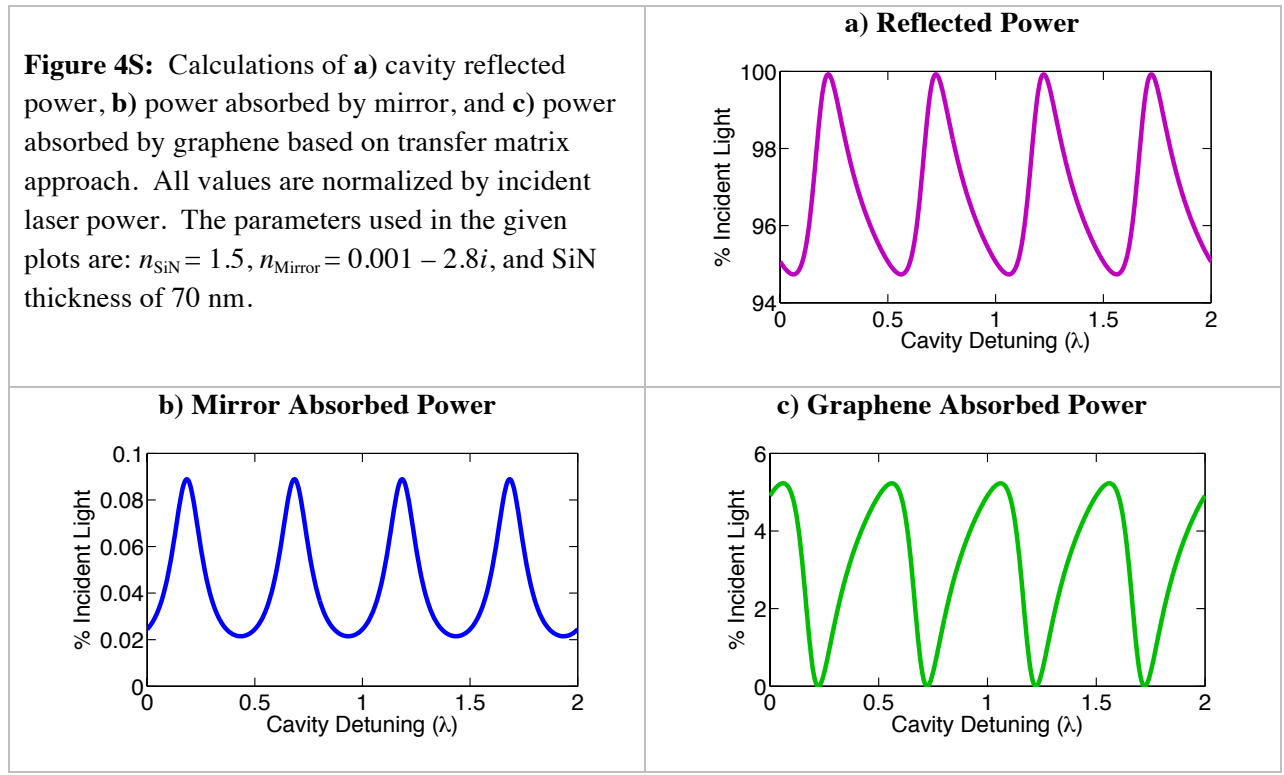

**Figure 4S:** Calculations of **a)** cavity reflected power, **b)** power absorbed by mirror, and **c)** power absorbed by graphene based on transfer matrix approach. All values are normalized by incident laser power. The parameters used in the given plots are: $n_{SiN} = 1.5$, $n_{Mirror} = 0.001 - 2.8i$, and SiN thickness of 70 nm.

## 5. Fitting of Cavity Detuning Data

With the calculations described above for the power in our optical cavity, a fitting model was generated for our resonant frequency vs. cavity detuning data. The resonant frequency shifts were taken to scale negatively with the optical power absorbed by the graphene layer – consistent with a tensile stress change caused by thermal expansion of the bilayer membrane. To aid in generating realistic optical parameters for our fit (refractive indices of the $Si_3N_4$ and the mirror), the amplitudes of our optically detected data were utilized. Nodal positions in the optical data should correspond to cavity detunings at which the gradient of the reflected light vanishes ($dR/dz=0$). Assuming small membrane deflections (relative to the optical wavelength), the signal amplitude scales linearly with the magnitude of this gradient, as in Figure 5S. The signal amplitude, however, is also affected by optical enhancement of the device Q, so only nodal positions are truly reliable.

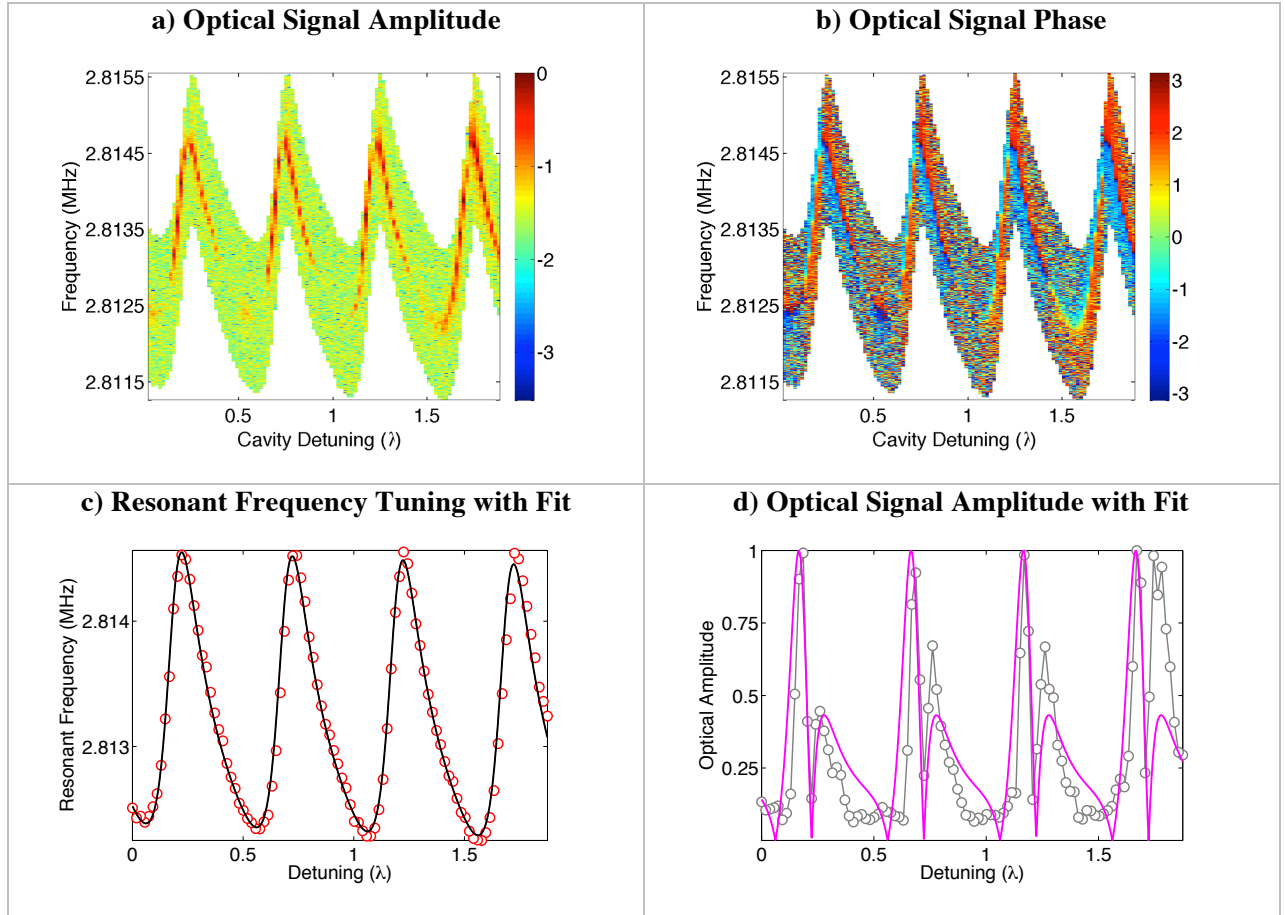

**Figure 5S: a)** Optically detected signal at 0.195 mW incident laser power. Color scale denotes log10 of normalized signal amplitude. **b)** Phase of optical data in (a). Nodes in optical signal are accompanied by 180° phase shifts. **c)** Electrically detected resonant frequency vs. cavity detuning. Fit scales negatively with optical power absorbed by graphene. Other fit parameters include a sloping background to account for capacitive softening of the resonator as the membrane is approached by the electrical gate. **d)** Normalized amplitude of optical signal vs. cavity detuning. Fit scales with absolute value of the gradient in the cavity's reflected optical power. Note the agreement in nodal positions between the data and fit.

## 6. Fitting of Damping Shift with Cavity Detuning

For low finesse cavities, photothermal forces dominate the optical feedback. The effective change in resonant frequency and damping due to a photothermal feedback force are given by [6,7]

$$\Gamma_{eff} = \Gamma(1 + Q\frac{\omega\tau}{1+\omega^2\tau^2}\frac{\nabla F}{K}), \quad \omega_{eff}^2 = \omega^2(1 - \frac{1}{1+\omega^2\tau^2}\frac{\nabla F}{K}) \quad \text{(Eq 5S)}$$

$\nabla F$ is the photothermal spring constant and is proportional to the absorbed power in the cavity ($\nabla F \propto dP_a(z)/dz$). K is the spring constant of the resonator and $\tau = L^2(\rho_{SiN}C_{SiN}t_{SiN} + \rho_g C_g t_g)/4(\kappa_{SiN}t_{SiN} + \kappa_g t_g)$ is the relaxation time[6–8] associated with the heat flow. We find that $\omega\tau = 2000$ for our resonator ($\omega$ = 17.6 MHz, L = 100 µm). Here, we used $\rho_{SiN}$= 3000 kg/m³, $k_{SiN}$ = 30 W/mK, $t_{SiN}$ =60 nm, $C_{SiN}$ = 700 J/Kg.K for silicon nitride, $\rho_g$= 2330 kg/m³, $k_g$ = 3000 W/mK, $t_g$ = 0.33 nm, $C_g$ = 750 J/Kg.K for graphene. Such a large time constant results in negligible shift in the resonant frequency of the resonator and hence observed frequency shifts can be attributed to static absorption-induced stress rather than the optomechanical forces. However, observed changes in the mechanical damping can still be attributed to the absorption dependent back action, the nature of which is still to be understood. Using the same optical parameters as Figures 4S & 5S and nodal positions in the optically detected data, this model produces a fit to the damping as shown in Figure 6S A. As can be seen in the figure, the data displays systematic deviations from the photothermal fit model. This is particularly true at high damping, where the data and model appear to be out of phase. Interestingly, the damping data seems to be 180° out of phase with the frequency tuning data (Figure 5S C). For this reason, we have also considered a damping variation model that scales with the graphene-absorbed power (Figure 6S B). While this fit seems in phase with the damping data near the maxima, it does not match the experiment well at low and intermediate values. A damping model that is a sum of these two contributions has also been considered (Figure 6S C), but still does not agree with experiment.

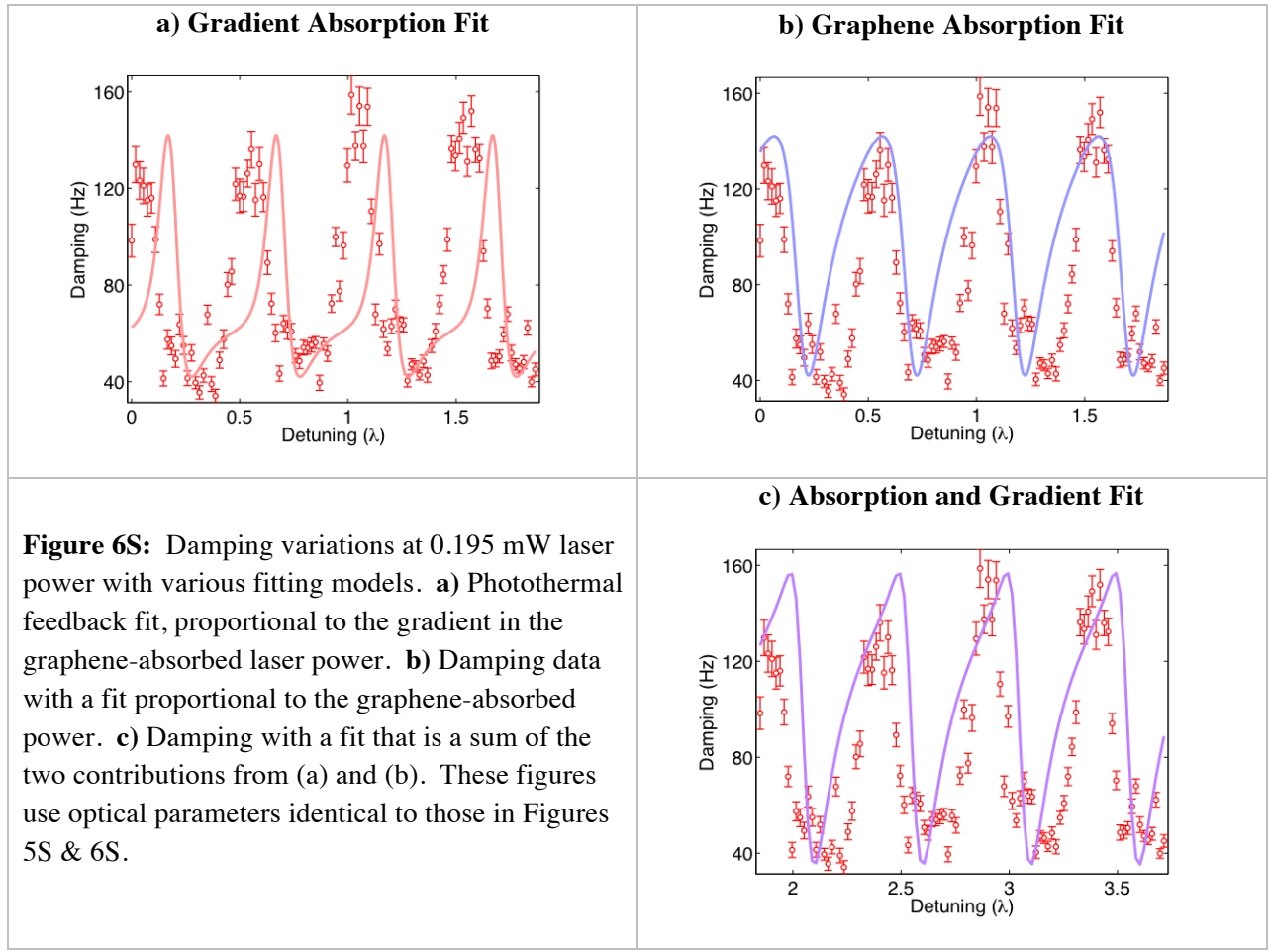

**Figure 6S:** Damping variations at 0.195 mW laser power with various fitting models. **a)** Photothermal feedback fit, proportional to the gradient in the graphene-absorbed laser power. **b)** Damping data with a fit proportional to the graphene-absorbed power. **c)** Damping with a fit that is a sum of the two contributions from (a) and (b). These figures use optical parameters identical to those in Figures 5S & 6S.

The complications in modeling the measured damping fits can arise from several sources. For extremely sharp resonances, for example, limited data point sampling near the peak frequency may affect the measured damping. It should be noted that the errorbars appearing in Figure 6S are based only on the goodness of the nonlinear least squares Lorentizan fits, and are not representative of the overall damping uncertainties.

Despite these issues in modeling the damping, the presence of a reproducible, optically-induced feedback force in our system is unmistakable. This effect on the damping is also seen in the amplitude of our electrically measured signal, as shown in Figure 7S.

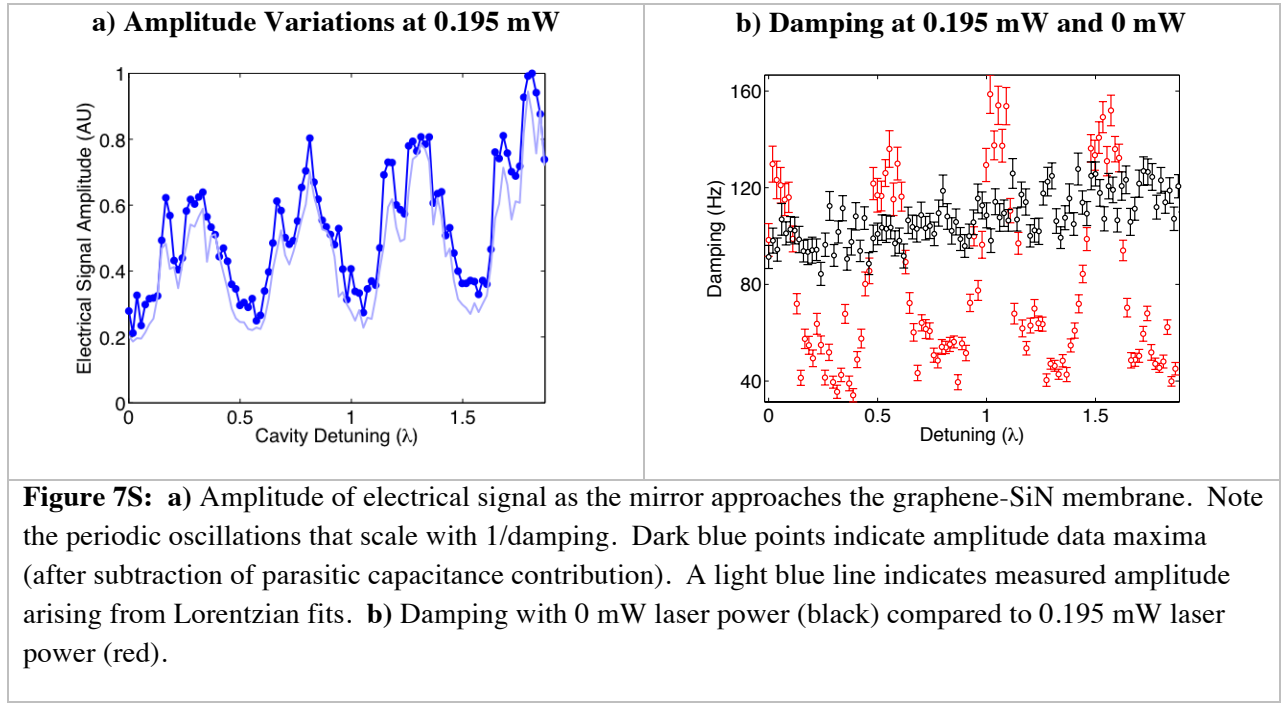

**Figure 7S: a)** Amplitude of electrical signal as the mirror approaches the graphene-SiN membrane. Note the periodic oscillations that scale with 1/damping. Dark blue points indicate amplitude data maxima (after subtraction of parasitic capacitance contribution). A light blue line indicates measured amplitude arising from Lorentzian fits. **b)** Damping with 0 mW laser power (black) compared to 0.195 mW laser power (red).

## 7. 300 um Device Response

All of the results shown thus far have been for a 100 μm graphene-on-silicon-nitride membrane. Figure 8S shows the optically detected response of a 300 μm graphene-on-silicon-nitride membrane of similar geometry (and identical thickness) to the 100 μm device depicted in Figure 1. The high Q is in line with that expected of a 300 μm SiN membrane of this tension. This suggests that (aside from optomechanics) the graphene has a minimal effect on the overall device mechanics.

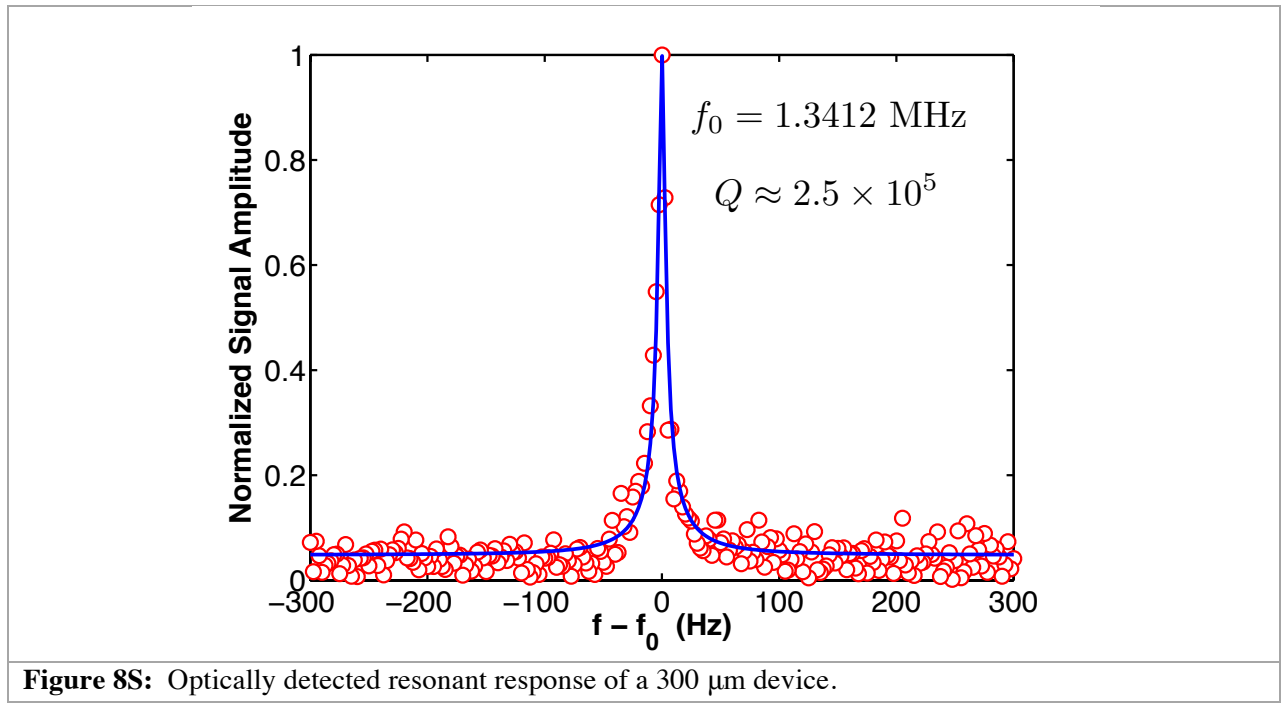

**Figure 8S:** Optically detected resonant response of a 300 μm device.